\documentclass[a4paper]{jpconf}
\usepackage{graphicx}
\usepackage{amsmath}
\usepackage{booktabs}
\usepackage{caption}
\usepackage{enumitem}
\usepackage{xcolor}
\begin{document}
\title{Improving science yield for NASA \textit{Swift} with automated planning technologies}

\author{Aaron Tohuvavohu}

\address{\textit{Swift} Mission Operations Center \\ Department of Astronomy and Astrophysics, The Pennsylvania State University\\ University Park, PA 16802}

\ead{aaronb@swift.psu.edu}

\begin{abstract}
The \textit{Swift} Gamma-Ray Burst Explorer is a uniquely capable mission, with three on-board instruments and rapid slewing capabilities. It serves as a fast-response satellite observatory for everything from gravitational-wave counterpart searches to cometary science. Swift averages 125 different observations per day, and is consistently over-subscribed, responding to about one-hundred Target of Oportunity (ToO) requests per month from the general astrophysics community, as well as co-pointing and follow-up agreements with many other observatories. Since launch in 2004, the demands put on the spacecraft have grown consistently in terms of number and type of targets as well as schedule complexity. To facilitate this growth, various scheduling tools and helper technologies have been built by the Swift team to continue improving the scientific yield of the Swift mission. However, these tools have been used only to assist humans in exploring the local pareto surface and for fixing constraint violations. Because of the computational complexity of the scheduling task, no automation tool has been able to produce a plan of equal or higher quality than that produced by a well-trained human, given the necessary time constraints. In this proceeding we formalize the Swift Scheduling Problem as a dynamic fuzzy Constraint Satisfaction Problem (DF-CSP) and explore the global solution space. We detail here several approaches towards achieving the goal of surpassing human quality schedules using classical optimization and algorithmic techniques, as well as machine learning and recurrent neural network (RNN) methods. We then briefly discuss the increased scientific yield and benefit to the wider astrophysics community that would result from the further development and adoption of these technologies.
\end{abstract}

\section{Introduction}
The scheduling of astronomical observations has been widely studied, for both space and ground based observatories. While there now exist a large number of robotic ground based telescopes, \textit{Swift} was the first semi-autonomous satellite observatory. Fast slewing and autonomous slew path determination and constraint checking were key design features of \textit{Swift} to allow it to achieve its mission goal of detecting, localizing, and rapidly settling on short Gamma-Ray-Bursts (GRBs) before they fade \cite{swift}. Due to these capabilities, \textit{Swift} was able to detect the first short GRB afterglow, and \textit{Swift} science was the major contributor to the contemporary consensus that the progenitor of short GRBs are compact object mergers (NS-NS or NS-BH). \textit{Swift} is a panchromatic mission with three instruments: the Ultraviolet/Optical Telescope, X-ray Telescope, and Burst Alert Telescope (UVOT, XRT, BAT) that are sensitive in the ultraviolet/optical, soft/hard X-ray, and gamma-ray regimes, with an energy range (non-continuous) spanning from ~2 eV up to 500 KeV. Due to its multi-wavelength capabilities, fast slewing, and rapid response time \textit{Swift} has become an enormously popular observatory for many different types of astronomical targets. While \textit{Swift} is nominally a GRB mission, over 80\% of observing time is now spent on non-GRB science targets. These include everything from tracking and observing comets in the solar system, to exoplanet science, to high-energy astrophysical sources at cosmological distances. Along with normal targets, \textit{Swift} also engages in large scale tiling campaigns, made possible by its rapid slewing and low re-pointing overhead. \textit{Swift} currently performs weekly tiling of the Small Magellanic Cloud \cite{jamie} and Galactic Bulge \cite{SBS}, as well as rapid-response tiling in response to astrophysical neutrino detections from ANTARES \cite{swiftantares} and IceCube \cite{swiftice} and gravitational wave triggers from the LIGO/VIRGO Collaboration \cite{evans}.

\begin{minipage}{\textwidth}
\begin{minipage}[t]{0.49\textwidth}
\vspace{0pt}
\flushleft
\includegraphics[width=3in]{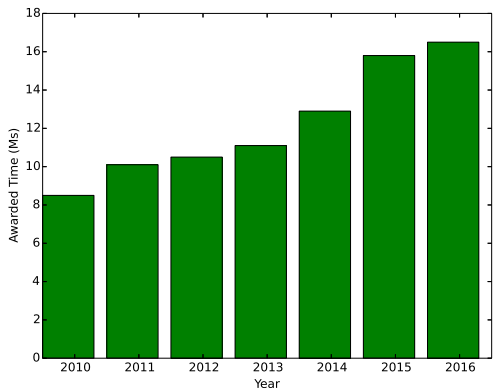}
\captionof{figure}{Amount of time awarded, in megaseconds, through the \textit{Swift} Target of Opportunity (TOO) program, from 2010-2016.}
\end{minipage}
\hspace{1cm}
\begin{minipage}[t]{0.40\textwidth}
\vspace{15pt}
\centering
\begin{tabular}{c c }
\toprule
\textbf{Year} & \textbf{Avg. \# Obs./day}\\
\midrule
2005 & 74 \\
2010 & 85  \\
2016 & 112 \\
\bottomrule
\end{tabular}
\captionof{table}{Average number of observations-per-day over the lifetime of the mission. In 2017 there were some days with $> 400$ individual observations.}
\end{minipage}
\end{minipage}

\vspace{\baselineskip}
The unique capabilities of the spacecraft, and the ever evolving nature of \textit{Swift} operations, have put unique demands on the operations team. Along with, and because of, \textit{Swift}'s abilities, it also has the most complex, dense, and dynamic observing schedules of any satellite observatory. The complex scheduling goals (science targets) are accompanied by a host of very particular constraints, both astronautical and scientific, that these schedules must meet (see \cite{passivecooling} for an example). In this proceeding we give a very brief overview of recent work performed with the goal of automating the creation of these sets of observing commands called Pre-Planned Science Timelines (PPSTs). Until now PPSTs have been built by (well trained) human Science Planners who were found to have a better understanding and command of the myriad constraints, and built markedly safer, more efficient, and higher quality PPSTs (with the assistance of a set of visualization tools) than any automation software available. However, building a PPST for one 24-hour period takes a human between 5-7 hours on average. In order to take maximal advantage of \textit{Swift}'s fast-response capabilities, the goal is to build an automated system that can build PPSTs at greater than human-level quality in under 30 minutes by using state-of-the-art optimization and modeling techniques.

This will allow the \textit{Swift} team to significantly increase the efficiency of the observatory and take maximal advantage of its fast-response capabilities when responding to ToO requests, while preserving as much pre-planned science as possible and maintaining spacecraft health and safety. This work will have a particularly large impact on \textit{Swift}'s search for electromagnetic (EM) counterparts to gravitational-wave triggers as in \cite{evans} and other rapid follow-up projects.

In section 2 we provide a precise formulation of the Swift Scheduling Problem (SSP), the solution to which is a PPST. In section 3 we motivate our specific approach through the analysis of large sets of simulated plans. In Section 4 we briefly motivate and outline some recent work on constraint modeling. In Section 5 we conclude with next steps and the impact of implementation of the proposed system. Detailed analysis and results regarding constraint modeling/satisfaction will be reported elsewhere.

\section{Formulation}
We state the SSP as a dynamic fuzzy Constraint Satisfaction Problem (DF-CSP). We adopt this approach as opposed to regular CSP because the SSP is over-constrained. That is, any complete assignment of the CSP violates some constraint. The fuzzy CSP scheme allows us to extend the problem to constraints that are not \textit{crisp}; constraints are assigned satisfaction degrees on the unit interval. We must also allow for new high priority ToO requests or transient detections (GRBs, Galactic transients, etc.). In this case, computing a completely new solution to the F-CSP is possible but has the distinct disadvantages of inefficiency and instability of the successive solutions, as described in \cite{reuse}. So we adopt the full DF-CSP framework as a method of reusing the old solution when constraints are added or updated, by making local changes to the relevant portion of the schedule.

In general, a CSP problem is defined by a set of variables $\mathcal{X}=\{X_i\}_{i=1}^n$ that take values on their associated finite domains $\mathcal{D}=\{D_i\}_{i=1}^n$ under a set of constraints $\mathcal{C}=\{C_i\}_{i=1}^r$. We make it \textit{fuzzy}, following the formalization in \cite{fuzzyaxioms}, by constructing $\mathcal{C}$ as a set of fuzzy constraints $R_i$ with the membership function $\mu_{R_{i}}$ indicating to what extent some label $\nu$ satisfies the $R_i$:
\begin{equation}\label{eq1}
\mathcal{C}=\Bigg \{ R|\mu_{R_i} :\prod_{x_{i}\in var(R_i)}D_{i}\to[0,1]\Bigg \}.
\end{equation}
So $\mu_{R_i}(\nu)$ gives the degree of satisfaction of the relevant constraint. The ability to treat crisp constraints is preserved by the extreme cases of $\mu_R(\nu)= 0 \mbox{ or } 1.$ However, $\mu_{R_i}$ gives only the \textit{local} degree of satisfaction; the degree to which the label $\nu$ satisfies the particular constraint. To judge whether a given $\nu$ is a solution to the full DF-CSP we need a measure of \textit{global} constraint satisfaction:
\begin{equation}\label{eq2}
\alpha(\nu)=\oplus \Big \{\mu_{R}(\nu_{var(R)})|R \in \mathcal{C} \Big\}
\end{equation}
where $\oplus$ is an aggregation operator on the unit interval. With $\alpha(\nu)$ one can now define a solution to the DF-CSP as a compound label $\nu_X$ for which $\alpha(\nu_X)\geq \alpha_0,$ where $\alpha_0$ is called the \textit{solution threshold,} which we determine based on need for optimality and run-time.

We also attach priorities $p$ to each of the constraints, which allows some constraints to be treated as more or less important than others. This allows one to be confident that, while some fuzzy constraints can be violated, no label $\nu_X$ can be found as a solution which \textit{completely} violates the most important constraints. This property matches the real-world need to preserve certain constraints that are critical to spacecraft health and safety. Adding priorities changes the specific definitions given in Eq. \ref{eq1} and Eq. \ref{eq2} slightly, a full formulation of the prioritized F-CSP can be found in \cite{fuzzyaxioms}. Either way, we have a formalization of our problem, and a clear solution to aim for: a label $\nu_X$ whose global constraint satisfaction $\alpha(\nu_X)$ passes the defined threshold $\alpha_0$.

All of the constraints, and associated cost function modeling by which we motivate the choice of priorities, can not be explicitly defined here, but we list most of the relevant hard and soft constraints below, with the most difficult to manage in red:

\begin{minipage}{\textwidth}
\begin{minipage}[t]{0.49\textwidth}

\center{Hard Constraints}
\begin{itemize}[leftmargin=*]
\item Slew resources capped.
\item {\color{red}Slew path constraints.}
\item Passive maintenance of XRT temperature between -55 C and -65 C.
\item {\color{red}Reaction wheel momentum limit.}
\item Roll constrained by solar-panel pointing and star tracker catalog.
\end{itemize}
\end{minipage}
\hspace{.20cm}
\begin{minipage}[t]{0.44\textwidth}
\center{Soft Constraints}
\begin{itemize}
\item {\color{red} Maximize observing/slew time ratio.}
\item Maximize number of science targets observed.
\item Minimize \# of UVOT filter wheel rotations.
\item Maximize time spent observing targets $>$ 8 hours from the sun.
\end{itemize}
\end{minipage}
\end{minipage}

\section{Approach}
Of course, one desires a strictly optimal solution to the scheduling problem, but combinatorial runtime is unacceptable: optimal scheduling for Swift is at least NP-hard. As the spacecraft has no consumables, a sub-optimal (but still highly optimized) plan that satisfies the various constraints and science priorities is acceptable. Early attempts set the solution threshold $\alpha_0$ quite high. However, it was found that these extremely optimized plans were too fragile. That is, the dynamic constraints were balanced so precisely that any disturbance (by ToO upload or GRB detection) would often break a hard constraint. Based on an analysis of simulation results, the author has good reason to believe that the effect of adding a constructed metric for `robustness' of a PPST to the list of soft constraints will be similar to that of a `good-enough' plan found by a heuristic scheduler. To wit, we motivate here the use of a multi-strategy scheduling system that is comprised of a heuristics-informed seed scheduler, guided by a generational genetic algorithm \cite{generation} which performs a greedy search, a gap-filler using a heuristic-biased stochastic sampling approach as in \cite{HBSS}, and a final-pass constraint fixer based on the contention measure \cite{contention}. 

Large sets of planning simulations were performed for various planning periods over many seasons to allow for diversity in the set of ephemerides. The results of these simulations confirmed the experiential wisdom of the Planners, that there is ``no one way to build a plan." That is, it was rigorously demonstrated that for any set of science targets that reasonably resembled the historical distribution of priorities for a given planning period, there exist multiple unique tours which fulfill the science priorities for the period, are safe for the spacecraft to perform (do not violate any hard constraints), and are reasonably similar in optimization score (soft constraints). Further it was shown that this holds for all relevant ephemerides, with a few marginal cases that can generally be alleviated with the assistance of a heuristic long-range planner.

While multiple unique acceptable solution classes exist, they were found to share general properties, with regard to target placement, that distinguish them from the majority of unacceptable tours. These properties are highly dependent on the details of the ephemeris for the given planning period. From the large simulation database, a set of meta-heuristics were derived for various classes of ephemerides. These meta-heuristics can be used to determine the appropriate scheduling strategy and partial search algorithm (local heuristic) for a given planning period. Tests of this approach have shown that once localized in a `good' regime of the solution space by the meta-heuristic informed seed schedule, it is almost always possible to quickly converge to an acceptable, and sometimes locally optimal, plan.

In the analysis of the large class of simulated tours, it was found that the success of a given tour is significantly more dependent on the uncertainty in certain constraints than others. It was further demonstrated that an increased understanding and better model of those constraints can have outsize impacts on the success and optimization of a given PPST. In particular, the rate of convergence to an optimized plan can be significantly hampered by getting `stuck' in unacceptable regions of the solution space because of an incomplete model of the momentum constraint. As such, a renewed effort at better understanding some of the constraints is valuable. A brief description of these efforts is given in Section 4.

\section{Constraint modeling}
In the course of investigating the solution space to the SSP, the critical constraint of preventing the saturation of the gyroscopic momentum wheels was found to have both the highest uncertainty in its modeling and the highest computational cost associated with that uncertainty. This motivates the construction of a better model. Constraint modeling for \textit{Swift} is almost uniquely suited to a machine learning approach, as the spacecraft has performed $>$ 350,000 individual observations so far in its mission. There is plenty of data to train on. In the investigation of the momentum constraint, previously unknown secular phenomena were discovered that make model building more difficult. However, over training periods of $\sim$ 10 days a recurrent neural network (RNN) with 4 hidden layers and ~50 nodes has shown promising results. This work will continue and results will appear in a future publication.
\section{Next steps and impact}
In this proceeding we formalized the SSP and motivated a particular approach to its solution; what remains is to finish building these planning algorithms into the \textit{Swift} planning infrastructure. The ability to quickly construct high quality PPSTs in response to new ToOs or transient events will allow \textit{Swift} to observe these targets in an optimal manner while minimizing the impact on pre-planned science priorities. Further, it will reduce inefficiency, continuous staffing needs, and pressure on members of the \textit{Swift} Science Operations Team. Finally, these operational improvements will significantly benefit multi-messenger and localization projects, and ensure that Swift is able to maximally contribute to the new age of gravitational wave and
multi-messenger astronomy.
\ack
This work was performed at the Swift Mission Operations Center at Penn State under support from the NASA Swift contract NAS5-00136. The author thanks his colleagues on the Swift FOT for their help in better understanding the spacecraft health \& safety constraints, Jeffrey Gropp for ongoing work relevant to constraint modeling and automated plan creation, and Alex Ledger for useful discussion pertinent to algorithmic optimization.

\section*{References}
\bibliography{iopart-num}

\providecommand{\newblock}{}
\begin{thebibliography}{10}
\expandafter\ifx\csname url\endcsname\relax
  \def\url#1{{\tt #1}}\fi
\expandafter\ifx\csname urlprefix\endcsname\relax\def\urlprefix{URL }\fi
\providecommand{\eprint}[2][]{\url{#2}}

\bibitem{swift}
Gehrels N {\em et~al.\/} 2004 {\em Astrophys. J.\/} {\bf 611} 1005--1020

\bibitem{jamie}
Kennea J~A {\em et~al.\/} 2018, in prep

\bibitem{SBS}
{Shaw} A~W {\em et~al.\/} 2017 ({\em AAS/High Energy Astrophysics Division\/}
  vol~16) p 400.01

\bibitem{swiftantares}
Adrian-Martinez S {\em et~al.\/} 2016 {\em JCAP\/} {\bf 2} 062
  (\textit{Preprint} \eprint{astro-ph/1508.01180})

\bibitem{swiftice}
Evans P~A {\em et~al.\/} 2015 {\em MNRAS\/} {\bf 448} 2210--2223
  (\textit{Preprint} \eprint{astro-ph/1501.04435})

\bibitem{evans}
Evans P~A {\em et~al.\/} 2017 {\em Science\/} (\textit{Preprint}
  \eprint{astro-ph/1710.05437})

\bibitem{passivecooling}
{Kennea} J~A {\em et~al.\/} 2005 {\em UV, X-Ray, and Gamma-Ray Space
  Instrumentation for Astronomy XIV\/} ({\em Proceedings of SPIE\/} vol 5898)
  pp 341--351

\bibitem{reuse}
Verfaillie G and Schiex T 1994 {\em Proceedings of the Twelfth National
  Conference on Artificial Intelligence (Vol. 1)\/} AAAI '94 pp 307--312

\bibitem{fuzzyaxioms}
Luo X, man Lee J~H, fung Leung H and Jennings N~R 2003 {\em Fuzzy Sets and
  Systems\/} {\bf 136} 151 -- 188

\bibitem{generation}
Ortiz-Bayliss J~C, Moreno-Scott J and Terashima-Marín H 2014  {\bf 512}
  315--327

\bibitem{HBSS}
Bresina J~L 1996 {\em Proceedings of the Thirteenth National Conference on
  Artificial Intelligence - Volume 1\/} AAAI'96 pp 271--278

\bibitem{contention}
Fraser S 2012 {\em Adaptive optimal telescope scheduling\/} Ph.D. thesis
  Liverpool John Moores University

\end{thebibliography}

\end{document}